\documentstyle[epsf]{metigm}
\begin{document}

\def\dim#1{\mbox{\,#1}}
\def\figname#1{metigm.#1}

\def\imageext{s}
\def\Mach{{\cal M}}
\def\hide#1{}

\title[Metal enrichment of the IGM]%
{Metal enrichment of the intergalactic medium}
\author[Gnedin]{Nickolay Y.\ Gnedin\\
University of California, Berkeley Astronomy Department,
Berkeley, CA 94720\\
e-mail: \sl gnedin@astron.berkeley.edu;
http://astron.berkeley.edu/\~{}gnedin}

\maketitle

\begin{abstract}
I demonstrate by means of high resolution cosmological simulations,
which include modelling of a two-phase interstellar medium,
that the dominant mechanism for transporting heavy elements from
the proto-galaxies into the IGM is the
merger mechanism as discovered by Gnedin \& Ostriker. Direct
ejection of the interstellar gas by supernovae plays
only a minor role in transporting metals into the IGM: for a realistic
cosmological scenario only a small fraction of all metals in the IGM
is delivered by the supernova-driven winds,
while most
of all metals in the IGM are transported by the merger mechanism.
As the result, the metallicity distribution in the IGM 
is highly inhomogeneous, in agreement with studies of the QSO metal
absorption systems, and the predicted metallicity distribution of Lyman-alpha 
absorbers as a function of their
column density is in excellent agreement with the observational data.
\end{abstract}

\begin{keywords}
cosmology: theory -- cosmology: large-scale structure of universe --
galaxies: formation -- galaxies: intergalactic medium -- intergalactic
medium: evolution -- hydrodynamics -- methods: numerical --
quasars: absorption lines
\end{keywords}

%
%

\section{Introduction}

It is now well established that the Lyman-alpha forest absorbers, 
thought to mostly consist of density peaks in the low density 
intergalactic medium
away from (proto-)galaxies (Cen et al.\ 1994; Zhang et al.\ 1995; 
Petitjean, Mucket, \& Kates 1995; Hernquist et al.\ 1996; 
Miralda-Escud\'{e} et al.\ 1996; Wadsley \& Bond 1996; Bi \& Davidsen 1996;
Hui, Gnedin, \& Zhang 1996), contain a measurable, about 1\% of solar,
amount of heavy elements (``metals'') at a redshift as high as three 
and even beyond. 
An early indication to this extent
by Meyer \& York (1987) is now confirmed quantitatively by 
new high-resolution Keck data (Womble, Sargent, \& Lyons 1995; Songaila
\& Cowie 1996), who found that as a column density of the Lyman-alpha 
absorbers decreases, smaller and smaller fraction of all absorbers contain
a measurable amount of heavy elements, indicating highly inhomogeneous
distribution of metallicities in the intergalactic medium (IGM). What is
the origin of those metals and their inhomogeneous distribution?

Within the boundaries of standard astrophysics, there is only one ``factory''
that produces metals - stars. Stars lose mass via winds and explode as
supernovae, polluting the interstellar medium with metals. However, from a
cosmological point of view, there is a long way from the interstellar
medium, which has a cosmological overdensity $\delta$
($\delta\equiv\rho/\bar\rho-1$) of about $10^4\div10^5$,
to the intergalactic medium with overdensities of about $10$ and below;
any mechanism that transports metals from the proto-galaxies into
the IGM should therefore be powerful enough to overcome the gravitational
force associated with enormous overdensities of 
(proto-)galaxies.\footnote{I avoid using the term ``galaxies'', which has
a well established meaning, since I will discuss properties of 
self-gravitating objects
at a range of redshifts, including high redshifts $z\sim10$, 
when no galaxies existed.}

For a long time since the pioneering works by Couchman \& Rees (1986) and 
Dekel \& Silk (1986; see also Miralda-Escud\'{e} \& Rees 1997 for a recent 
review) it has been thought that the supernovae blowing out gas from
the Population
III objects would do the job. However, recent numerical three-dimensional
simulations (Ostriker \& Gnedin 1996; Gnedin \& Ostriker 1996) as well
as semi-analytical results (Haiman, Rees, \& Loeb 1996)
convincingly demonstrated that Pop III objects, formed via molecular 
hydrogen cooling, are so few (mostly because the molecular hydrogen is 
so fragile), that they are unable to produce the metallicity
in excess of about $10^{-5}$ solar. 

Nevertheless, supernovae can still
disrupt normal, Pop II, proto-galaxies and pollute the IGM with metals, as
argued, for example, by Miralda-Escud\'{e} \& Rees (1997). If it were the
only mechanism for transporting metals from the deep potential wells of
proto-galaxies into the diffuse IGM, it would be the final answer to the
question: ``What is the metal transport mechanism in the IGM?''. However,
as was discovered by Gnedin \& Ostriker (1996),
there exists another metal transport mechanism, which is 
based solely on 
gravitational hierarchical 
clustering. Namely, when two proto-galaxies merge with a
sufficiently large relative velocity in a close to head-on collision, 
dark matter halos and stellar components of two halos
penetrate through each other,
whereas gas components of both objects collide in a gigantic shock. The
pressure force arising from this interaction is comparable to the
gravitation force from the dark matter and baryons of two objects. Therefore,
the collisionless component (the dark matter and stars) and the collisional
component (the interstellar gas) 
experience forces that significantly differ in their
directions and amplitudes, and, as the result, both components may have
different spatial distributions in the process of merging. Because the
gas constitutes only a small fraction of the total mass of a proto-galaxy,
it often happens during those violent mergers that some fraction of the gas
(typically a few percent), 
displaced relative to the dark matter, becomes gravitationally unbound and 
gets ejected from the proto-galaxy into the IGM. This gas, which was a part
of the interstellar medium in the past, and is heavily enriched with metals,
gets mixed up with the surrounding IGM, polluting it with heavy elements.
As the result, the metal distribution
in the IGM becomes highly inhomogeneous:
while regions around the proto-galaxies are 
enriched with heavy elements, the centers of large voids, well removed
from the places where merging occurs, still contain the pristine material
(an example of how this mechanism works is shown in Gnedin \& Ostriker 1996
and will not be presented in this paper). 

One can imagine other ways to transport metals from the
proto-galaxies to the IGM.
For example, a fraction of young stars can be lost during a merger, and 
going off later as supernovae, may enrich the IGM. I consider this process
a part of merging mechanism, and will not distinguish it from the case when
the metal enriched gas rather than stars was lost in the merger. After all,
since stars
are subject to the same (purely gravitational) force as the 
dark matter is, and the gas is not, it is much easier to unbound a substantial
fraction of the total gas component of two merging proto-galaxies rather 
than of their stellar component.

The whole process of the metal enrichment of the IGM begins in stars: stars
make metals in the first place in the gas with the cosmic overdensity of
$\delta\sim10^{30}$; blowing up as supernovae, they enrich the ISM 
($\delta\sim10^{5}$) with metals; metals from the ISM get transported into
the IGM ($\delta\la10$) either by the very same supernovae or by the merging
mechanism as explained above. It is therefore the purpose of this paper to
investigate the relative importance of those two metal transport mechanisms
within the framework of a reasonable cosmological model.

Since the transition between the ISM and the IGM is somewhat uncertain,
and in order to avoid introducing another, ``intermediate'', component
for the cosmic gas, I will call ``the ISM'' the gas that is gravitationally
bound to a particular proto-galaxy, and ``the IGM'' the rest of the
cosmic gas (i.e.\ gas, which is 
not bound to a particular self-gravitating object).

Several attempts have been made before to include effects of supernovae into
cosmological hydrodynamic simulations 
(c.f.\ Cen \& Ostriker 1992; Navarro \& White 1993; Katz, Weinberg, 
\& Hernquist 1996). However,
all previous works attempted to model the ISM as a {\it single\/}-phase
medium, simply injecting thermal and kinetic energy of supernovae into
the cosmic gas. This approach is strongly dependent on the numerical
resolution of a simulation. For a low resolution simulation, injection
of thermal energy may produce a significant effect. However, as the resolution
of a simulation increases, and higher and higher density regions are
being resolved, the cooling time in the gas becomes smaller and smaller,
and any amount of the thermal energy that is directly injected into the cosmic
gas, is radiated away before it has had a change to influence the dynamics
of the gas. In reality however, the thermal energy gets injected in supernova
remnant bubbles (the second phase in the gas), that are too hot and do not cool
as efficiently.

On the other hand, the kinetic energy directly injected into the gas can
propagate a large distance from the source of injection, because the
unsufficient resolution of a simulation prevents one to resolve the
snow-plow phase of the supernova remnant evolution 
(Cioffi, McKee, \& Bertschinger 1988; McKee 1990),
during which most of the kinetic energy of the supernova ejecta is
absorbed. Thus, the effect of the finite numerical resolution can lead
to both underestimating and overestimating the effect of the supernovae
on the dynamics of the ISM, depending on the resolution of a simulation
and details of the implementation of the supernova feedback. Because the
resolution of a high-resolution Lagrangian simulation is not uniform
across the simulation volume, and varies from point to point, it is
extremely difficult to disentangle the physical effect of supernovae
feedback from numerical artifacts produced by the finite resolution of a 
simulation.

It is therefore logical to attempt to model the multi-phase ISM in order
to account properly for the supernovae feedback. Again, the finite resolution
of existing cosmological hydrodynamic simulations does not allow to resolve 
properly the evolution of a single supernova remnant, but the multi-phase
ISM can be incorporated in a cosmological simulation as ``sub-cell''
physics, similar to the way the star formation is included in simulations.
In other words, for a sufficiently large region, which contains many
supernovae, it is possible to account for the total effect, a population
of supernova remnants have (on average) on the dynamics of the interstellar 
gas (energy injection, extra pressure, reduction in the star formation rate 
etc). 

Simulations reported on in this paper attempt to do precisely that. The main
difference between this work and previous investigations is that a
{\it two\/}-phase medium is now allowed for in a simulation, i.e.\ at
{\it each point\/} in a simulation the cosmic gas has two temperatures
(the ``cold'' phase outside the supernova remnants and ``hot'' phase
inside the supernova remnants) and two densities (which can be expressed
as the total density and the volume fraction occupied by the hot phase).
The current implementation
explicitly assumes however that the population of supernova remnants
is moving together with the rest of the cosmic gas, i.e.\ the two
phases have the same velocity on a large ($\sim 1\dim{kpc}$) scale.
The velocity itself is however affected by the feedback from the
supernova remnants. The specific recipe for incorporating physics of 
the two-phase ISM into a cosmological hydrodynamic code is presented in
Appendix A.

Finally, a few words about interpreting results of a cosmological simulation.
Any ``sub-cell'', phenomenological description of a physical process
like supernova feedback or star formation inevitably introduces new parameters
that describe the physical process under consideration. It is currently
unrealistic (from the point of view of computational resources)
to perform a full-scale extensive parameter study, and thus it should be
kept in mind that {\it any\/} cosmological hydrodynamic simulation which
includes ``sub-cell'' physics (including simulations reported on in this 
paper) should be considered as a {\it plausible possibility\/} rather than
the {\it final answer\/}. There always remains a chance that the nature
picked up different values of parameters than those used in a simulation,
however plausible the latter seem to be.

\section{Method}

I have adopted a CDM+$\Lambda$ cosmological model as a framework for
these investigations. The qualitative results of my
simulation are applicable to the whole family of the CDM-type models,
and I emphasize those qualitative conclusions together with specific
quantitative results for the model I use. The cosmological parameters
are fixed as follows:
$$
\begin{array}{rcl}
        \Omega_0 & = & 0.35, \\
        \Omega_\Lambda & = & 0.65, \\
               h & = & 0.70, \\
        \Omega_b & = & 0.055, \\
\end{array}
$$
which is one of the cosmological models adopted by the Grand Challenge
Cosmological Consortium (GC$^3$) as the ``best-bet'' models
({\sl http://arcturus.mit.edu/GC3/IC\/}). This model
is similar to the model investigated by Gnedin \& Ostriker (1996) except
that $\Omega_b$ has a higher value, in concordance with the low primordial
abundance as inferred by Tytler \& Burles (1996).

Since the merging mechanism can only be studied by a numerical simulation,
I use the SLH-P$^3$M cosmological hydrodynamic code as described by Gnedin
(1995), Gnedin (1996), Gnedin \& Bertschinger (1996), and Gnedin \& Ostriker 
(1996). However, since the goal of this paper is to study the effects of
supernovae on the metal transport in the IGM, a proper numerical treatment
of the physical effect of supernovae has to be incorporated in the 
code. Appendix A explains how the physics of the
multi-phase interstellar medium
can be incorporated into a cosmological hydrodynamic code on scales large
compared to the size of one supernova bubble. Three parameters
control the evolution of the ISM in this regime: $E_{51}$, the average 
energy of a supernova measured in units of $10^{51}\dim{erg}$, $f_{\rm IMF}$,
a factor parametrizing the excess (lack) of metal production compared to
the Miller-Scalo IMF, 
and $v_{\rm term}$, the termination velocity of the supernova
shock at which the interior of the bubble is ionized. In order to minimally
span the parameter space and understand how supernovae effect the IGM on
different scales, six simulations have been performed. Each simulation
has $64^3$ dark matter particles, $64^3$ gas cells, and a number of stellar
particles which are being formed during a simulation. The softening parameter
for all simulations is set to $1/10$, corresponding to the dynamical range
of $640$. 
In addition to the new ingredient, the physics of the supernova, 
physical modelling in all
runs includes all components described in Gnedin (1996) and 
Gnedin \& Ostriker (1996), to which papers the reader is referred 
for more details about physical modelling.
Here I only note that I allow for the fully three-dimensional treatment
for the evolution of the dark matter and the two-phase cosmic gas, as well as
for the formation and destruction of molecular hydrogen and
all other standard physical processes for a gas of primeval composition,
following in detail the ionization and recombination of all species in
the ambient radiation field. The spatially and frequency dependent
radiation field, in turn, allows for sources
of radiation (quasars and massive stars), sinks (due to 
continuum opacities), self-shielding of the dense gas lumps, and 
cosmological effects. In regions which are cooling
and collapsing (beyond the numerical resolution of the code) 
I allow the formation of point-like ``stellar''
subunits, permitting them to release radiation and explode as supernovae.
Supernovae inject thermal and kinetic energy in the ISM (via the second
gaseous phase - the supernova remnants), and pollute the
surrounding gas with metals, which are in turn considered in the treatment 
of cooling.

Parameters of all runs are summarized in Table \ref{tableone}. 
\def\tableone{
\begin{table}
\label{tableone}
\caption{Numerical Parameters}
\medskip
\begin{tabular}{lccccc}
Run & Box size & Spatial res.\ & $E_{51}$ & $f_{\rm IMF}$ & $v_{\rm term}$\\ 
A1 & $1h^{-1}{\rm\,Mpc}$ & $1.6h^{-1}{\rm\,kpc}$ & 0 & 4 & NA \\
B1 & $1h^{-1}{\rm\,Mpc}$ & $1.6h^{-1}{\rm\,kpc}$ & 1 & 4 & $c_S$ \\
C1 & $1h^{-1}{\rm\,Mpc}$ & $1.6h^{-1}{\rm\,kpc}$ & 1 & 4 & $30\dim{km}/\dim{s}$\\
A3 & $3h^{-1}{\rm\,Mpc}$ & $4.7h^{-1}{\rm\,kpc}$ & 0 & 4 & NA \\
B3 & $3h^{-1}{\rm\,Mpc}$ & $4.7h^{-1}{\rm\,kpc}$ & 1 & 4 & $c_S$ \\
C3 & $3h^{-1}{\rm\,Mpc}$ & $4.7h^{-1}{\rm\,kpc}$ & 1 & 4 & $30\dim{km}/\dim{s}$\\
\end{tabular}
\end{table}
}
\tableone
In order to span a large enough range in mass scales, two sets of runs
are performed: the first set with the $1h^{-1}\dim{Mpc}$ box size and the
total mass resolution of $10^{5.6}h^{-1}{\rm M}_\odot$,
and the second set with the
$3h^{-1}\dim{Mpc}$ box size and the total mass resolution 
of $10^{7.0}h^{-1}{\rm M}_\odot$. 
The resolution in baryons is 
a factor of $0.055/0.35=0.16$ finer, and for runs with the $1h^{-1}\dim{Mpc}$ 
box size, it is $10^{4.8}h^{-1}{\rm M}_\odot$, which is about the Jeans
mass at $z\sim15$ for this model. Those runs, therefore, account for all
initial fluctuations that are present in the baryons, i.e.\ they do not miss
any small scale power in the initial conditions (but those simulations
do miss the small
scale power that is generated by the nonlinear dynamics at later times
on the scales below
the resolution limit).

Simulations within each set have had
exactly the same initial conditions; so that evolution of individual
objects can be followed under different physical assumptions.
Because it makes no sense to continue
a simulation of a finite size beyond the redshift when perturbations with
the wavelength of the order of the box size become nonlinear, runs with
$3h^{-1}\dim{Mpc}$ box size are stopped at the redshift $z=4$, 
and runs with 
$1h^{-1}\dim{Mpc}$ box size are not continued beyond $z=9$.

In each of two sets three different runs
have been performed: 
\begin{description}

\item[run A,] when no physics of the multiphase ISM (as
described in Appendix A) has been taken into account (manifested by the
parameter $E_{51}$ set to zero); this run does not
include effects of supernovae, and therefore only the merging mechanism
for metal transport operates in this case; this run is then used for
comparison with two other
runs where effects of supernovae are included, to assess the
relative importance of two metal transport mechanisms.

\item[run B,] which includes physics of the multiphase ISM; in this run
it is assumed that the
supernova shock is terminated at the sound speed of the ambient medium.

\item[run C,] which includes physics of the multiphase ISM; in this run
the supernova shock is terminated at the velocity of $30\dim{km}/\dim{s}$.
\end{description}
The latter choice is somewhat arbitrary; it approximately
maximizes the effect of supernova on heating of the low metallicity ISM.
As has been shown by Shull \& McKee (1979), the supernova shock in the
ISM with solar metallicity fails to ionize the gas at shock velocities
below $90\dim{km}/\dim{s}$. It is expected that for low metallicity 
gas at high redshift this value is somewhat lower (Shull \& Silk 1981),
and I somewhat arbitrary fix it to be $30\dim{km}/\dim{s}$, near the
value which maximizes the effect of supernovae. This choice also allows
for the possibility that the star formation occurs inside the dense
supernova swept-up shells, which are considered belonging to the
cold phase for the shock velocities below $30\dim{km}/\dim{s}$.

Finally, I choose a factor $f_{\rm IMF}=4$, which implies that I assume
a top-heavy IMF, with a factor of four higher metal production then in
the solar neighborhood. As argued by Silk (1996), this value is supported
by metal abundances in the clusters of galaxies if one assumes that all metals
came from cluster galaxies alone. This number is an extreme,
in a sense that if the clusters of galaxies contain additional stars which are
distributed
over the cluster (and which would be virtually unobservable),
then $f_{\rm IMF}$ should be less than 4. Both runs C1 and C3 should therefore
be considered as limiting cases when all possible parameters are pushed
to their extreme values in order to maximize the effect of supernovae.

\section{Results}

\subsection{The Metallicity of the IGM}

\def\capSF{
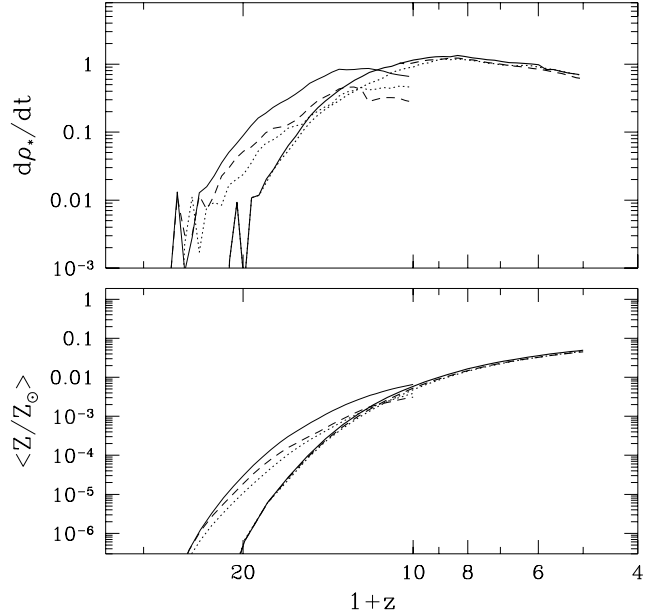
{\it Upper panel\/}: the
comoving star formation rate in units of $\dim{M}_\odot/\dim{yr}/(h^{-1}
\dim{Mpc})^3$ as a function of redshift for six runs: 
A1 ({\it thin solid line\/}),
B1 ({\it thin dotted line\/}),
C1 ({\it thin dashed line\/}),
A3 ({\it bold solid line\/}),
B3 ({\it bold dotted line\/}), and
C3 ({\it bold dashed line\/}).
{\it Lower panel\/}: the mass weighted mean metallicity of the cosmic gas
as a function of redshift; line markings are as on the upper panel.
}
\begin{figure}
\par\centerline{%
\epsfxsize=1.0\columnwidth\epsfbox{\figname{figSF.ps}}}%
\caption{\label{figSF}\capSF}
\end{figure}
As the cosmic structure develops under the influence of gravity, 
baryons condense
into bound objects, cool and eventually form stars, which end their lives
as supernovae and pollute the environment (i.e.\ ISM and IGM) with metals.
Figure \ref{figSF} shows in the upper panel the comoving
star formation rate in 
units of $\dim{M}_\odot/\dim{yr}/(h^{-1}\dim{Mpc})^3$ for all six runs:
the bold lines indicate runs A3-C3, and the thin lines
correspond to runs A1-C1. 
The solid lines
mark runs A, the dotted lines mark runs B, and the dashed lines mark runs C.
In all figures below line markings are the same unless specified otherwise.
Note that the star formation declines at late epochs due to the final
size of a simulation box (and, as the result, lack of large-scale
power to drive the gravitational collapse further).

The lower panel of the same figure shows the mass-weighted average 
metallicity of the
cosmic gas (i.e.\ ISM and IGM together) as a function of redshift.
I would like to note here,
that simulations with different box sizes are reasonably consistent
with each other in a sense that runs with the $3h^{-1}\dim{Mpc}$ box size
(A3-C3) have similar star
formation rates and metallicities at $z=9$, when runs with the  
$1h^{-1}\dim{Mpc}$ box size (A1-C1) become unreliable because of the missing
large-scale power.

As can be seen, both the star formation rates and the 
average metallicities for
runs with the $1h^{-1}\dim{Mpc}$ box size are somewhat suppressed in cases
when supernova effects were taken into account (runs B1 and C1, thin dotted 
and dashed lines) as compared
to the case when no supernova effects were included (run A1, the
thin solid line).
It is the expected result, since as the volume fraction occupied by supernova
bubbles increases, a smaller and smaller fraction of the gas is available for
star formation, because the gas inside the bubbles is hot and is not able
to cool
fast enough. However, even on the smallest resolved scales
this effect is not dramatic:
in the most favored case (C1) both the star formation rate and the average
metallicity are decreased by $z=9$ by a factor of 2.5, but not by orders of
magnitude.

\def\capMF{
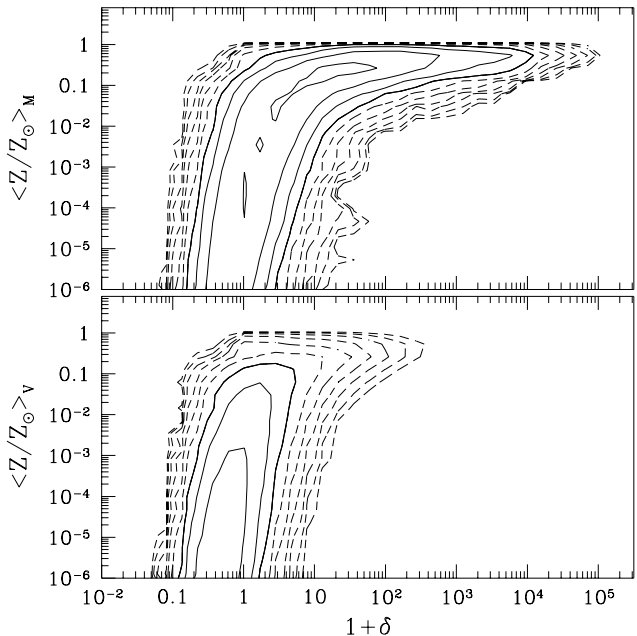
The mass- ({\it upper panel\/}) and volume- ({\it lower panel\/}) weighted
probability distribution to find a fluid element with given values of the
metallicity and the cosmic gas density for run A3 at $z=4$. 
The bold line shows the contour corresponding to the
probability $P(Z=\langle Z\rangle,\delta=0)/4$, solid contours mark the 
probability higher and the dashed
contours show the probability lower than the one marked with the 
bold contour; contour spacing is logarithmic
with the increment of $1/3$ dex.
}
\begin{figure}
\par\centerline{%
\epsfxsize=1.0\columnwidth\epsfbox{\figname{figMF.ps}}}%
\caption{\label{figMF}\capMF}
\end{figure}
Before I proceed further with more detailed comparison between the two main
mechanisms of metal transport in the IGM, it is instructive to discuss
the spatial distribution of the metallicity. Figure \ref{figMF} shows
the mass- (the upper panel) and the volume-weighted (the lower panel) 
probability distributions for a fluid element to have given values of
the metallicity and the gas density (as measured by the overdensity 
$\delta=\rho/\bar\rho-1$) from run A1 at $z=4$ (a similar figure
is presented in Gnedin \& Ostriker 1996).
The immediate conclusion is that the 
distribution is quite broad, with regions at about the cosmic mean density
having values of metallicity all the way from zero to about solar
(at $z=4$ !). The
distribution narrows down in the high density limit, because the high
density gas resides inside the proto-galaxies and is enriched with metals.
This high-density, metal-rich gas occupies only a small fraction of the
volume, so, if there is a substantial fraction of the gas with metallicities
as high as tenth of solar (at $z=4$ !), most of the volume is occupied
by the low metallicity gas. This wide distribution of metallicities is in
agreement
(at least qualitatively) with the recent conclusions by 
Rauch, Haehnelt, \& Steinmetz (1996) and by Hellsten et al.\ (1997),
who found that the observed scatter in the abundance ratios of the QSO
metal absorption systems is inconsistent with the assumption of
the uniform metallicity. The required scatter in the metallicity at a given
value of the cosmic density is about an order of magnitude, which is
consistent with Fig.\ \ref{figMF} for $\delta\ga10$.

I emphasize here that Fig.\ \ref{figMF} presents the total cosmic gas,
including both the ISM and the IGM. The gas with overdensities in excess
of $10^3\div10^{4}$ is apparently located inside the proto-galaxies, and is
identified with the ISM. The low density gas, $\delta\la10$, is clearly
a part of the IGM (and is, therefore, partly pristine), with the gas at
intermediate densities belonging to the ISM or the IGM, depending on
whether it is bound to a particular self-gravitating object or not. 

\def\capZD{
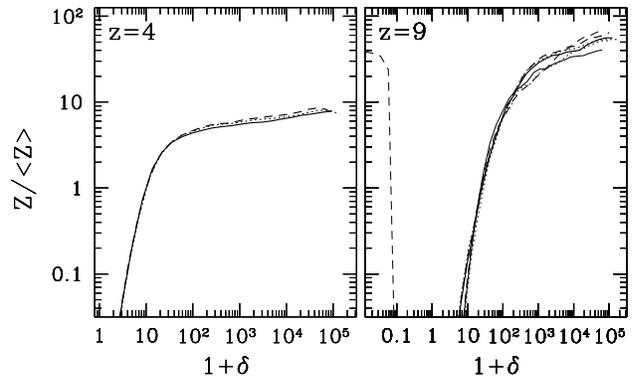
The average metallicity of the gas as a function of the cosmic gas 
density at $z=4$ ({\it left panel\/}) and $z=9$({\it right panel\/}).
Line markings as in Fig.\ \protect{\ref{figSF}} except for $z=4$ case where
no lines for runs A1-C1 are shown since those runs were stopped at a higher 
redshift. 
}
\begin{figure}
\par\centerline{%
\epsfxsize=1.0\columnwidth\epsfbox{\figname{figZD.ps}}}%
\caption{\label{figZD}\capZD}
\end{figure}
In order to compare the metallicity-density relationship for different
runs, I consider how the average metallicity 
changes with the gas density. For 
of all fluid elements having the overdensity $\delta$ within the
range from $\delta$ to $\delta+d\delta$, I compute their
metallicity $Z(\delta)$ as a function of $\delta$ and plot it
in Figure \ref{figZD} in units of the mass-weighted average metallicity
for all six runs (in the other words, the solid bold line in the left panel
is Fig.\ \ref{figMF}
collapsed along the vertical direction, and other lines are produced from
the metallicity-density distributions for other respective runs;
line markings are again as in Fig.\ \ref{figSF}). The left
panel shows runs (A3-C3) at $z=4$ (since runs A1-C1 were stopped at $z=9$),
and the right panel shows all six runs at $z=9$. 

The fact that all three lines almost coincide in the left panel
of Fig.\ \ref{figZD} demonstrates that at intermediate redshifts
($z\sim4$) and large scales ($\sim100\dim{kpc}\div1\dim{Mpc}$) 
supernovae play no role
in transporting metals into the IGM. It is the merging mechanism
that enriches the IGM with heavy elements and produces a highly
inhomogeneous distribution of metallicities. Even at higher redshifts,
and on smaller scales, supernovae play a minor role as a metal transport
mechanism in the IGM. However, in the right panel of Fig.\ \ref{figZD}
($z=9$) one can notice that in run C1 (the most favorable for the supernovae
effects) there is a fraction of gas sitting at very low densities
(about 0.01 of the mean) and having the metallicity 40 times above the
average (about 10\% solar). This would
be a signature of the supernova ejected gas, 
which would
be highly enriched with metals and would have low densities, since its
temperature is higher (about $10^5\dim{K}$) that the average temperature
of the IGM at this high redshift ($10^{3}\div10^{4}\dim{K}$). 
However, except for 
this very low density gas, differences in the metallicities 
between runs with supernovae (B,C)
and runs without supernovae (A) are not significant for the IGM with
the density higher than the cosmic mean.

\def\capZZ{
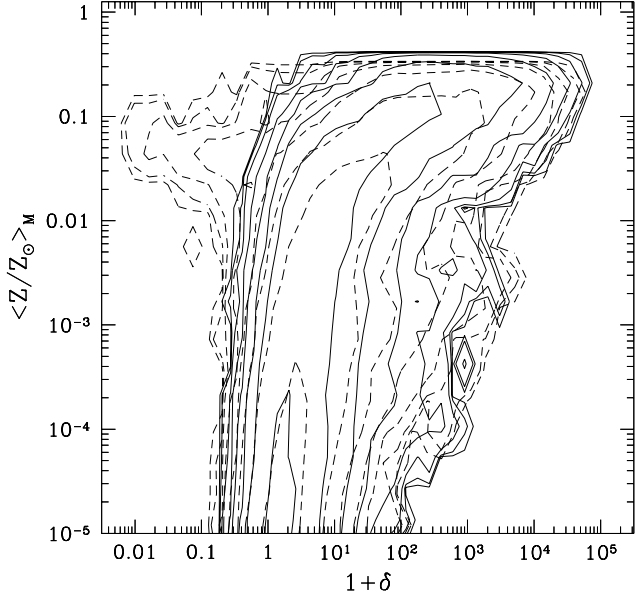
The mass-weighted
probability distribution to find a fluid element with given values of the
metallicity and the cosmic gas density for runs A1 
({\it solid lines\/}) and C1 ({\it dashed lines\/}) at $z=9$. 
The contour spacing
is as in Fig.\ \protect{\ref{figMF}}. The low-density - high-metallicity region
of the plane in run C1 (supernova blown gas) accounts for 0.03\% of the 
total gas mass and 0.6\% of the total volume.
}
\begin{figure}
\par\centerline{%
\epsfxsize=1.0\columnwidth\epsfbox{\figname{figZZ.ps}}}%
\caption{\label{figZZ}\capZZ}
\end{figure}
The difference between runs A1 (no supernovae) and C1 (most efficient
supernovae) is further demonstrated in
Figure \ref{figZZ}, which shows the mass-weighted 
metallicity-density distributions
for those two runs in a fashion similar to the upper panel of
Fig.\ \ref{figMF}, except
that two different line markings now correspond to two runs: the
solid lines show contours for runs A1, and the dashed lines show contours
for run C1. The two distributions are very similar except for a small
amount of the very low density - high metallicity gas which is 
present in run C1
but is absent in run A1. This gas is blown out of a proto-galaxy by the
supernova activity; its amount is insignificant: it occupies 0.6\% of
the total volume and amounts to 0.03\% of the total gas mass in the universe.
This gas is not necessarily the only gas that was ejected from the
proto-galaxies
by supernovae: there are slight differences in the metallicity-density 
distributions in runs A1 and C1 at higher densities, which are due to 
differences in the star formation rates and the supernova activity (but those
two can not be meaningfully separated). However, 
in the whole, the supernova mechanism plays a minor
role in transporting metals into the IGM, and by $z\sim4$ gas disruption
in violent mergers is responsible for delivering of more 
than 99\% of all heavy elements in the IGM.

One may ask a question: what is fraction of the enriched gas in the colliding 
high redshift objects which gets unbound and ejected into the IGM? 
It might not be appropriate to talk about a single value, but it is easy
to come up with an upper limit. Let me consider an
oversimplified picture of the universe, which consists of high-density
bound objects with the average metallicity $Z_H$, which contain $f_H$
of the total mass. The rest of the universe is in the IGM, with the
average metallicity $z_L$. I also assume that bound objects occupy a
negligible fraction of the volume. In this case, the volume weighted
averaged metallicity will be just $Z_L$. For the run A3 at $z=4$, this value is
$Z_L=6\times10^{-3}Z_\odot$. The mass weighted average metallicity in
this case is $\langle Z\rangle_M=Z_Hf_H+Z_Lf_L$. 
For run A3 at $z=4$ the respective values are $f_H=0.2$ and
$\langle Z\rangle_M=5\times10^{-2}Z_\odot$, which implies that the average
metallicity of bound objects $Z_H=0.23Z_\odot$. Since all
metals in the IGM come from the gas that gets unbound in mergers, the
fraction $f_{\rm unb}$ of this gas is
\[
	f_{\rm unb} = {Z_Lf_L\over Z_Hf_H+Z_Lf_L} = 10\%.
\]
This number is however an upper limit, since I have assumed that the bound
objects occupy no volume at all, because any clustering in the IGM will
decrease this bound as well, and because the limited resolution of 
my simulations underestimates the mass fraction $f_H$ locked in bound
objects. 

\hide{
Why are the supernovae so
inefficient in expelling the gas from proto-galaxies? 
It is instructive to make a simple estimate: let me consider a supernova
going off in the neutral metal poor ISM with the density of hydrogen
$n_H=1\dim{cm}^{-3}$ (which corresponds to a relatively modest overdensity
of 4000 at $z=9$). The supernova ejects $10^{51}\dim{erg}$ of mechanical
energy.
The expanding shell terminates at the sound speed of the ambient
medium. Using equations (\ref{tpds}) and (\ref{rpds})
it is easy to estimate that for the
ambient medium with the temperature of $10^4\dim{K}$ the supernova
bubble exists for only $2.6\times10^6$ years and 
terminates at the radius of $90\dim{pc}$; for the case
when the ambient medium has a temperature of $10^3\dim{K}$, the
corresponding life time is $1.3\times10^7$ years and the
radius is $150\dim{pc}$. Those distances are still
smaller than the virial radius of $330\dim{pc}$ of an object with
the baryonic mass of $3\times10^4{\rm M}_\odot$ (the Jeans mass) at $z=9$.
This implies that several supernovae have to go off {\it simultaneously\/}
(within a few million years; for comparison, the age of the Universe
at this moment is 500 million years) in order to expel a large fraction
of the gas
even from the least massive object. 
}

\hide{
Let me make a simple estimate. The Milky Way galaxy, with
$5\times10^{10}\dim{M}_\odot$ of baryons in the star forming disk, 
produces one supernovae every 30 years, at a rate 
\[
	R_{\rm MW}=
	{1 \dim{Supernovae}\over 5\times10^{10}\dim{M}_\odot
	\times 30\dim{years}}.
\]
In order to disrupt a $3\times10^4\dim{M}_\odot$
object at $z=9$, one would need $(330/150)^3\sim10$ supernovae
going off within $10^7$ years. This requires the rate
\[
	R_{\rm JM} = 
	{10 \dim{Supernovae}\over 3\times10^{4}\dim{M}_\odot
	\times 10^7\dim{year}} = 50 R_{\rm MW}.
\]
Because the low mass objects with
virial temperatures below about $10^{4}\dim{K}$ can cool only via molecular
hydrogen cooling or metal cooling, and because both processes are
inefficient due to the fragility of the molecular hydrogen and low
metallicity, it
is unlikely the supernovae rate in those objects can be 50 times higher
than the supernovae rate in the Milky Way galaxy. 
}

\hide{
For more massive objects ($M\la10^6\dim{M}_\odot$), whose virial temperature
reaches $10^4\dim{K}$, the estimate for the required supernova rate would
be even higher, because both the virial radius increases and the average
size of supernova remnants decreases with increase in the gas temperature.
}

\hide{
Finally, the above estimate is more likely a lower limit than a real
value, because it is not enough to expel the gas just beyond the virial
radius: the gas would still be in the collapsing flow
and would be accreted to the object in the next Hubble time. In order to
completely expel gas from a bound object, the gas should be ejected beyond
the turn-around radius, in which case the estimate above should be increased
by some factor (less than about 10).
}

More than ten years ago Dekel \& Silk (1986) analysed the effect of
supernovae on the gas ejection from high redshift proto-galaxies using the
analytical approach. They found that
supernovae can efficiently expel gas from low mass proto-galaxies 
provided that the
star formation is extremely rapid and is happening on a free-fall time
scale.
I point out here that there are
two important physical effects, which are included in the simulations
presented here,
but omitted in the Dekel \& Silk (1986) calculations.
First, the negative feedback
[eq.\ (\ref{negfeed})] limits the star formation
rate when the fraction of the volume occupied by supernova remnants becomes
large. This effects is ignored in Dekel \& Silk (1986) paper. Second,
Dekel \& Silk (1986) ignore cooling of the interstellar gas, which
leads to a substantial loss of energy injected by the supernova.
For example,
if all of the supernova energy was simply injected into the ISM gas 
without proper
account for the two-phase ISM and supernova remnant evolution, {\it all\/}
energy would be radiated away, producing negligible net effect.

One can therefore ask a question whether the present
work conflicts with the conclusions of Dekel \& Silk (1986). I would
like to emphasize here that this paper does not pretend to claim
that the supernovae do not eject gas from bound objects, and as I show
in the following section, they sometimes do it very efficiently. Thus,
the supernova ejection {\it is\/} a mechanism for metal transport into the
IGM. However, as I explained above, there is {\it another\/} metal
transport mechanism, the merger mechanism, and the simulations presented
in this paper support the conclusion that the merger mechanism is
more efficient. The major physical reason for that is that the merger
mechanism works irrespectively of the binding energy of an object and
of the star formation rate in the object, whereas
supernovae are able to disrupt the gas from a bound object only for
objects with sufficiently low masses and high star formation rates. 
More than that, 
in order to transfer metal
enriched gas into the IGM, it is required to expel the gas not just beyond
the virial radius of the dark matter halo, but beyond the turn-around radius;
otherwise, the expelled gas will fall back into the dark matter halo in a
fraction of a Hubble time. 

Therefore, it has to be kept in mind that the results presented in this 
paper do not
necessarily contradict the conclusions of Dekel \& Silk (1986), but 
a more careful comparison will have to be performed in order to
verify whether the simulations actually confirm the conclusions of
Dekel \& Silk (1986).

\subsection{Supernovae in the Proto-Galaxies}

\label{expel}

\def\capCC{
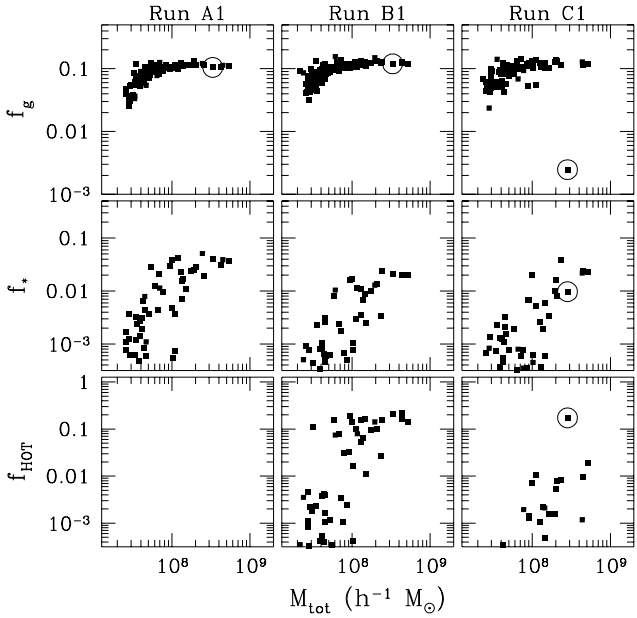
Properties of all 
bound objects containing more than 100 particles identified
at $z=9$ in runs A1 ({\it left column\/}), B1 ({\it middle column\/}), and 
C1 ({\it right column\/}). Shown in three rows as a function of the total
mass of an object are the gas fraction for
each object ({\it upper row\/}), the stellar fraction ({\it middle row\/}),
and the fraction of the hot gas within the supernova remnants 
({\it lower row\/}). The only object whose gas was
completely 
disrupted by the supernova activity in run C1, and the same objects in runs
A1 and B1, are marked with a circle in the upper row and in the right column.}
\begin{figure}
\par\centerline{%
\epsfxsize=1.0\columnwidth\epsfbox{\figname{figCC.ps}}}%
\caption{\label{figCC}\capCC}
\end{figure}
As I have demonstrated in the previous section, the
supernova-driven winds are less efficient in transporting metals in the IGM
than the merger mechanism.
Nevertheless, supernova disruption of proto-galaxies
does happen, as the difference between runs A1 and C1 is due to the gas
disrupted by the supernovae from the proto-galaxies. In order to understand
this process, I show in Figure \ref{figCC} properties of all well-resolved
(containing more than 100 particles of all kinds) bound objects identified
at $z=9$ in runs A1 (the left column), B1 (the middle column), and 
C1 (the right column). Three rows show the gas fraction, the stellar
fraction, and the fraction of the hot gas within the supernova remnants
for each object versus its total mass (the lower left panel is empty
because the supernova effects are not included in run A1).
The fraction of the hot gas never exceeds about 30\% since it is
averaged over the whole object, including the outer regions where there is
no star formation (and thus the supernova activity); 
in most of the objects it reaches close to 
100\% in the center, where star formation takes place. In run C1 fractions
of the hot gas are smaller that the respective fractions in run B1, because
the supernova remnants are assumed to terminate earlier in run C1. However,
the total effect of supernovae is actually larger in run C1 because each
supernova remnant excerts much larger pressure on the ambient ISM and
ejects much more thermal energy than a respective supernova remnant in
run B1.

In all three columns distributions are similar, except that
in the third column
(run C1) there is a single object whose gas fraction is very small, and
whose gas was apparently expelled by supernovae. This object and respective
objects in runs A1 and B1 are marked by a circle (since all three runs A1-C1
started with the same initial conditions, there is a one-to-one correspondence
between the bound objects in all three runs).
One can also notice at
least three objects in the same panel lying somewhat below the main 
distribution. Those objects have also lost some of their gas due to the
supernova activity, but their gas fraction is reduced by about a factor 
of 2, and not by two orders of magnitude. This would imply that the
total disruption of a bound object by supernovae is a relatively rare
event, but the supernova-driven winds are much more common.

\def\capCT{
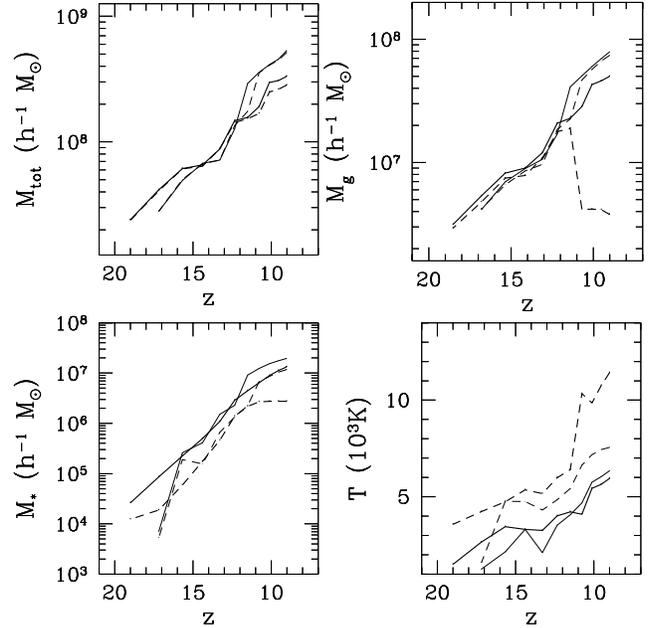
Evolutionary tracks for objects from runs A1 and C1 marked with circles
in Fig.\ \protect{\ref{figCC}} ({\it bold lines\/}) together with
evolutionary tracks for the most massive objects from the same runs
({\it thin lines\/}). Line marking is as before: solid lines show
run A1, and dashed lines show runs C1. The gas disruption occurred
at $z=11$. Four panels show the total mass ({\it upper left panel\/}),
the stellar mass ({\it lower left panel\/}), the gas mass
({\it upper right panel\/}), and the average gas temperature
({\it lower right panel\/}) as a function of redshift.
}
\begin{figure}
\par\centerline{%
\epsfxsize=1.0\columnwidth\epsfbox{\figname{figCT.ps}}}%
\caption{\label{figCT}\capCT}
\end{figure}
The evolution of the disrupted object (the bold lines)
is presented in Figure \ref{figCT}
together with the evolution of the most massive object (the thin lines)
in a simulation. Four panels show the total mass (the upper left panel),
the stellar mass (the lower left panel), the gas mass
(the upper right panel), and the average gas temperature
(the lower right panel) as a function of redshift. The disrupted object is
shown by the bold dashed line. The act of disruption, which is manifested
by a sharp drop in the gas mass and an equally sharp rise in the
temperature occurred at $z=11$. It was preceded by a steep rise in
the object's stellar mass, which is a signature of the burst of star 
formation.
For comparison, the most massive object (the thin lines) has similar
behavior of both the total and the stellar mass until the moment of
disruption, with two exceptions: its stellar mass is rising slightly
less steeply
just before z=11, and its total mass continues to rise, whereas the total
mass of the disrupted object rose insignificantly from $z=12.5$ to
$z=11$. It is therefore plausible to assume that the disruption of the
gas contents of the object is a result of a rare coincidence of
two events happening almost
simultaneously: a burst of star formation and a sharp decline in the
accretion rate. It is not very likely that those two processes happen
simultaneously, and, therefore, the total disruption of the gas component of
a proto-galaxy is a rare event.
The more quantitative discussion of the process
of gas disruption is however beyond the framework of this paper.

\def\capPA{
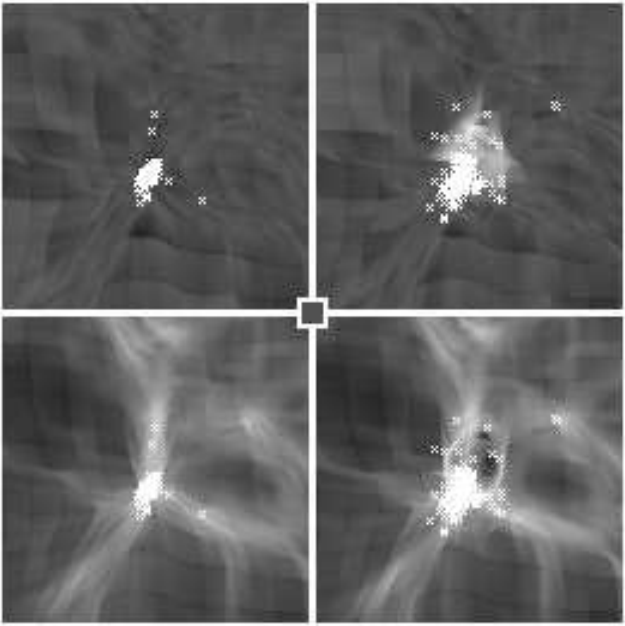
A slice of the gas density ({\it lower row\/}) and the gas temperature
({\it upper row\/}) of comoving dimensions
$200h^{-1}\dim{kpc}\times200h^{-1}\dim{kpc}\times15h^{-1}\dim{kpc}$
around the disrupted object at $z=11$ for runs A1 ({\it left column\/})
and C1 ({\it right column\/}). Stars (i.e.\ SLH stellar particles) are shown
by white symbols. The small square at the center mark the mean cosmic density
for the density panels and $T=10^{3}\dim{K}$ for the temperature panels.
The color version of this plot is available online as explained in the
Appendix B.
}
\begin{figure}
\par\centerline{%
\epsfxsize=1.0\columnwidth\epsfbox{\figname{figPA\imageext.ps}}}%
\caption{\label{figPA}\capPA}
\end{figure}
\def\capPB{
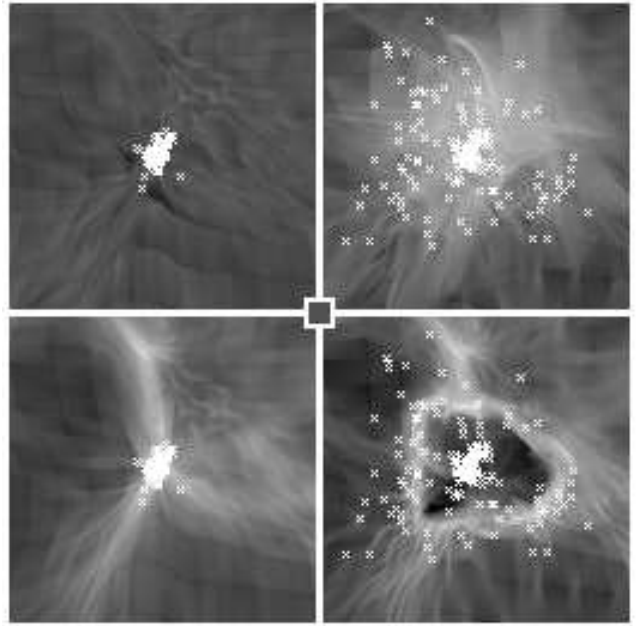
The same as Fig.\ \protect{\ref{figPA}} except at $z=9$.
}
\begin{figure}
\par\centerline{%
\epsfxsize=1.0\columnwidth\epsfbox{\figname{figPB\imageext.ps}}}%
\caption{\label{figPB}\capPB}
\end{figure}
\def\capPC{
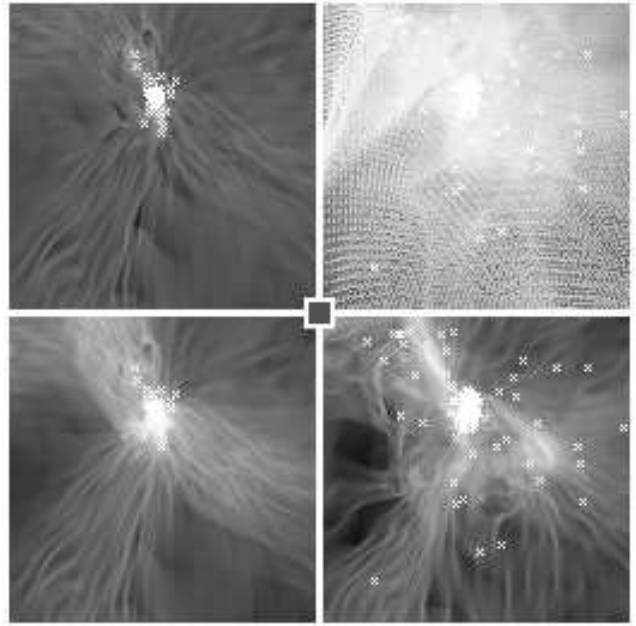
The same as Fig.\ \protect{\ref{figPA}} except at $z=7$. The upper right
panel is of somewhat poorer quality than other panels because of 
incomplete rebinning.
}
\begin{figure}
\par\centerline{%
\epsfxsize=1.0\columnwidth\epsfbox{\figname{figPC\imageext.ps}}}%
\caption{\label{figPC}\capPC}
\end{figure}
Figures \ref{figPA}-\ref{figPC} serve to illustrate the process of
gas disruption. These figures show a neighborhood
of the object marked with a circle in Fig.\ \ref{figCC}. The scale of each
panel is $200h^{-1}\dim{kpc}\times200h^{-1}\dim{kpc}$ in comoving units.
The upper panels show the 
gas temperature, and the lower panels show the gas density
for the same object in runs A1 (the left panels) and C1 (the right panels). 
Fig.\ \ref{figPA} shows the object at the moment of disruption, $z=11$, and
Fig.\ \ref{figPB} shows the same comoving volume at $z=9$. In order to 
illustrate the further evolution of this objects, runs A1 and C1 are
continued until $z=7$, with 
Fig.\ \ref{figPC} showing the same comoving volume at $z=7$. Stars are shown
on the same plots with white stellar symbols. 

The process of disruption started at $z=11$ with the ejection of the gas in the
``north-east'' direction (assuming that north is on the top of the figure),
because the high-density filaments prevented the shock wave to expand
into the north and east directions. The low density - high temperature
region formed inside the shock wave. At a later time, $z=9$ 
(Fig.\ \ref{figPB}), the shock wave overcame the confining pressure of 
filaments and expanded in all directions; a very low gas density region
formed around the proto-galaxy, which now contains almost entirely stars and
the dark matter (not shown). The expanding shock wave snowplowed a large
amount of external material, and secondary shock waves were propagating
in all directions inside the expanding bubble. As it expanded and
snowplowed the surrounding gas, the shock wave also managed to disrupt
a small fraction of stars from the object; stars interact with the other
matter only gravitationally, but since the gas constitutes some 15\% of the
total mass of the object (for a considered cosmological model), the mass of the
expanding shell is not negligible, and some stars were disrupted from the
proto-galaxy by the gravitational pull of the expanding shell. By $z=7$ 
(Fig.\ \ref{figPC}), the expanding shell dissolved and mixed with the 
surrounding IGM, heating the
gas to much higher temperature than it would otherwise had.

For comparison,
the left columns (run A1 without supernovae) show nothing even close
to this spectacular activity.

An MPEG video of this process as well as color versions of
Fig.\ \ref{figPA}-\ref{figPC} are available online as explained in
Appendix B.

\subsection{Metal Abundances in the Lyman-alpha Forest}

\def\capZL{
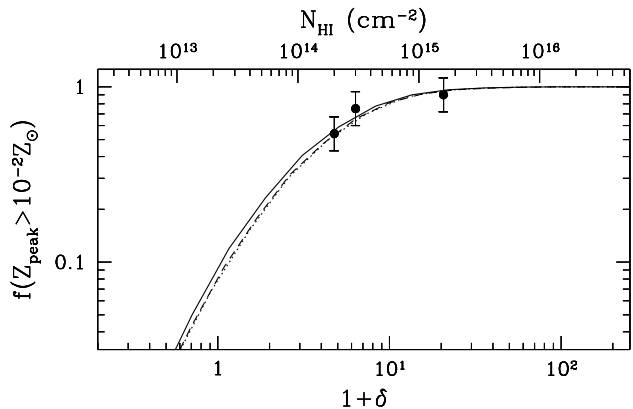
The fraction of all peaks in the gas density which have the metallicity
above the limiting value $Z_{\rm lim}=10^{-2}Z_\odot$ as a function of 
the cosmic gas density for runs 
A1 ({\it solid lines\/}),
B1 ({\it dotted lines\/}), and
C1 ({\it dashed lines\/}) as extrapolated to $z=3$. The upper scale shows
the average column density of a Lyman-alpha forest line arising from a 
respective density peak (Hui et al.\ 1996). Points with errorbars are
observational data from Womble et al.\ (1995) and Songaila \& Cowie (1996)
(the data point from Womble et al.\ (1995) is corrected by 17\% for 
$Z_{\rm lim}=10^{-2}Z_\odot$ from their value of 
$Z_{\rm lim}=10^{-2.3}Z_\odot$ using the dependence of the fraction of all 
peaks above $Z_{\rm lim}$ from run A1).
}
\begin{figure}
\par\centerline{%
\epsfxsize=1.0\columnwidth\epsfbox{\figname{figZL.ps}}}%
\caption{\label{figZL}\capZL}
\end{figure}
While both the theoretical modelling and the observations of the metallicity
of the Lyman-alpha Forest are still in the developing stages, it is
still instructive to make a rough comparison of the results of
simulations with the observational data from Womble et al.\ (1995) and
Songaila \& Cowie (1996). 
As was demonstrated by Hui et al.\ (1996), most of the Lyman-alpha
lines with column densities below about $10^{16}\dim{cm}^{-2}$
are arising from the peaks in the line-of-sight distribution
of the cosmic gas density. 
Figure \ref{figZL} shows by the bold lines
the fraction of all one-dimensional 
density peaks (i.e.\ peaks along any one of three directions)
with the metallicities in excess
of 1\% solar as a function of the peak density at $z=3$. Three almost
coinciding lines correspond to three runs: A3 (the solid line),
B3 (the dotted line), and C3 (the dashed line). Since all three
runs were stopped at $z=4$, I have extrapolated the metallicity distributions
from those three runs to $z=3$.

Since the density
of a peak strongly correlates with the column density of a Lyman-alpha
line, it is possible to establish a one-to-one correspondence between
the peak density and the column density of the absorption line. Respective
column densities are presented on the upper scale in Fig.\ \ref{figZL}
for a considered cosmological model (the correspondence between the
peak density and the column density of a Lyman-alpha line depends
on cosmological parameters). Solid dots with errorbars are data from
Womble et al.\ (1995) (the leftmost point) and from Songaila \&
Cowie (1996). Since Womble et al.\ (1995) quote results for the
limiting metallicity of $10^{-2.3}Z_\odot$, I have corrected their
value for the limiting metallicity of 1\% solar using the dependence
of $f(Z>Z_{\rm lim})$ on $Z_{\rm lim}$
as derived from run A3 (by 17\% percent). 
A somewhat arbitrary 25\% errorbar was added to each data point. 
 
One can note that the agreement between the observations and the simulations
is pretty good, taking into account that no fitting has been performed
to achieve this agreement. However, this agreement should only be
considered as encouraging, but by no means definite agreement between
the theoretical predictions and the observations. Since several theoretical
uncertainties such as limited resolution of my simulations, scatter in
the column density -- peak density relation etc, are not included in 
the analysis, it remains to be seen if more accurate simulations
(including generating synthetic Lyman-alpha spectra and analyzing
them by fitting Voigt profiles) will confirm this agreement.

It also remains to be seen how sensitive this test is. Since the
merging mechanism is the dominant metal transport mechanism in the IGM,
and the amount of merging strongly depends on the amplitude of
density fluctuations, one may expect that cosmological
models with different amplitudes of density fluctuations on small
scales (several hundred kpc) will have substantially different
predictions for the fraction of Lyman-alpha lines above a certain
metallicity as a function of a column density.

\section{Conclusions}

I have demonstrated by means of high resolution cosmological simulations
that the dominant mechanism for transporting heavy elements from
the high density cosmic gas (ISM) into the low density cosmic gas (IGM) is the
merger mechanism as discovered by Gnedin \& Ostriker (1996). Direct
ejection of the interstellar gas by supernovae plays
only a minor role in transporting metals into the IGM: for a realistic
cosmological scenario (i.e.\ the one that has a reasonable amount
of power on all relevant scales from the {\it COBE\/} scale all the way to
the scale of about $100\dim{kpc}$ probed by the low column density 
Lyman-alpha forest), and at the most
favorable conditions, only a small fraction of all metals is directly
expelled from proto-galaxies by supernovae, while most
of all metals in the IGM are delivered by the merger mechanism. It does
not mean that supernovae do not expel gas from proto-galaxies, and
examples in Section 3.2 show that they indeed do. However, the main
conclusion of this paper is that the other, merger mechanism is more
important in transporting metals into the IGM (in particularly, because
it is {\it not\/} a subject to the negative feedback process), and
ignoring it in computing the metal transport rate in the IGM is
inadequate. 

As the result, the metallicity distribution in the IGM 
is highly inhomogeneous, in agreement with conclusions by 
Rauch et al.\ (1996) and Hellsten et al.\ (1996). 

The metallicity distribution of Lyman-alpha absorbers as a function of their
column density is in excellent 
agreement (subject to the unknown uncertainty due to the limitations
of theoretical modelling) with the observational data
from Womble et al.\ (1995) and Songaila \& Cowie (1996).

Apparently, the effect of the supernovae depends on the star formation rate.
One may therefore that the star formation rate can be underestimated in
the simulations, thus leading to a erroneous conclusion that supernovae
are not efficient in expelling metals from the ISM. While details of
the specific implementation of star formation in the simulations reported on
here can be found in Gnedin (1996), I note here that the algorithm adopted
in the simulation assumes that the star formation occurs on the dynamical
time in the rapidly cooling gas at a rate, which, if being continued until
$z=0$, would substantially overpredict the stellar mass density in the
universe at the current epoch. Thus, the star formation rate in the
simulations, as all other
relevant parameters, is pushed to the reasonable extreme in order to
maximize the effect of supernovae. And still, even in the most favorable
conditions, supernovae do not contribute significantly to the metal
transport rate into the IGM at high redshift ($z\la4$).

This conclusion is only relevant in the range of redshifts simulated in this
paper, and for the class of cosmological models whose properties are
similar to the LCDM model considered here. The current work makes 
no predictions about the relative importance of supernova driven winds at
lower redshifts, where higher star formation rate and lower density
of the cosmic gas in the virialised objects may lead to enhanced role
of supernova driven outflows.

As a side note, I mention here that since the merging mechanism works
irrespectively of the binding energy of a collapsed object, merging
in clusters of galaxies at the current epoch also ejects metals in the
IGM. Thus, I can make a prediction that, if the conclusions of this
paper are indeed correct, the mean metallicity of the IGM at the current epoch
can easily be as high as several percent, or even 10\% of 
solar.\footnote{Assuming that 10\% of the gas lost in mergers of clusters of
galaxies, which contain the intracluster gas with the $1/3$ solar metallicity,
leads to the conclusion that a 3\% solar metallicity in the IGM is produced
just from clusters of galaxies at the current epoch.}

I am very grateful to Chris McKee for his numerous discussions with me
and explanations of 
the physics of the multi-phase interstellar medium. I thank M.\ Rees, 
J.\ Silk, J.\ Ostriker, A.\ Dekel, S.\ Zepf, M.\ Steinmetz, and 
J.\ Miralda-Escud\'{e} 
for valuable comments. The manuscript has been substantially improved
by comments and suggestion by the anonymous referee.
This work was supported by the 
UC Berkeley grant 1-443839-07427.
Simulations were performed on the NCSA Power Challenge Array under 
the grant AST-960015N and on the NCSA Origin2000 mini-supercomputer 
under the grant
AST-970006N.

\appendix

\section{Incorporating the Multi-Phase Interstellar Medium into a Cosmological
Simulation}

\def\Nu{\mbox{\LARGE$\nu$}}

As has been shown by Cioffi et al.\ (1988), a supernova
remnant radius $R_S$ in the pressure driven snowplow stage (when radiative
losses set in) is given by the following expression:
\begin{equation}
	R_S = 25.7\dim{pc}
	{E_{51}^{31/98}\over \xi^{5/98} n_0^{18/49} v_{100}^{3/7}},
	\label{rpds}
\end{equation}
where $E_{51}$ is the energy of the supernova remnant in units of 
$10^{51}\dim{erg}$, $n_0$ is the ambient hydrogen density in units of
$1\dim{cm}^{-3}$, $v_{100}$ is the shock velocity $v_S$ in units of 
$100\dim{km}/\dim{s}$, and
\[
	\xi = \min\left(Z/Z_\odot,3\times10^{-3}n_0^{1/2}E_{51}^{1/4}
	\right),
\]
as shown by Cioffi \& Shull (1991). This radius is reached at the time
$t_S$ after the explosion,
\begin{equation}
	t_S = 7.57\times10^4\dim{ys}
	{E_{51}^{31/98}\over \xi^{5/98} n_0^{18/49} v_{100}^{10/7}}.
	\label{tpds}
\end{equation}
For the very low density ambient medium $n_0<n_{\rm crit}$, 
when radiative there are no losses, where
\[
	n_{\rm crit} = 2.8\times10^{-3}\dim{cm}^{-3}
	{v_{100}^7\over\xi^{3/2}E_{51}^{1/2}},
\]
the supernova remnant radius is given by the following expression:
\begin{equation}
	R_S = 27.8\dim{pc}
	{E_{51}^{1/3}\over n_0^{1/3} v_{100}^{2/3}}.
	\label{rnopds}
\end{equation}

The volume fraction $f_{\rm hot}$ of the ISM occupied by the 
supernova remnants is
determined by the {\it porosity\/} $Q$,
\[
	f_{\rm hot} = 1 - \exp(-Q),
\]
where (McKee 1990; Silk 1996)
\begin{equation}
	Q = \Nu {1\over m_{\rm SN}} {d\rho_*\over dt}.
	\label{qdef}
\end{equation}
Here $\Nu$ is the 4-volume of the supernova remnant in the space-time,
$d\rho_*/dt$ is the rate of star formation, and $m_{\rm SN}$ is the
mass in stars formed per supernova.
The 4-volume of the supernova remnant is given by the following
expression:
\begin{equation}
	\Nu = q {4\pi\over3} R_S^3 t_S,
	\label{nudef}
\end{equation}
where a dimensionless parameter $q$ depends on the 
time-evolution of the supernova
remnant.

I now assume that the remnant evolution terminates at the shock velocity
$v_{\rm term}$. If $v_{term}=\beta c_S$, where $c_S$ is the sound speed of
the ambient ISM, and $\beta$ is a parameter of order unity (Cioffi et 
al.\ 1988), then the 4-volume of the supernova remnant until the moment
of maximum expansion is (McKee 1990)
\[
	\Nu_1 = {1\over 3\eta+1} {4\pi\over3} R_S^3 t_S,
\]
where $\eta=0.3$ is the power-law index of the expansion law,
$R_S\sim t_S^\eta$. Comparing to equation (\ref{nudef}), I get
$q_1=1/(3\eta+1)$. However, in this case one must also account for the
contraction phase, which gives $q_2=1/(4\eta)$ (McKee 1990). The total
$q$ in this case (which I will call Case A hereafter) is
\[
	q_A = {1\over 3\eta+1} + {1\over4\eta}.
\]
The volume of the ISM occupied by the supernova remnants contains hot
ionized gas, and, therefore, no stars can form inside the supernova bubbles.
However, as was shown by Shull \& McKee (1979), a supernova remnant shock
fails to ionize the ambient medium with the solar metallicity 
at the shock velocities below about
$90\dim{km}/\dim{s}$. In this case the interstellar gas snowplowed by the
supernova shock with a lower velocity will not be ionized and may form
stars even if it located inside the supernova remnant. In order to account
for this possibility, I consider $v_{term}$ as a free parameter, not
necessarily close to the sound speed of the ambient ISM. Obviously,
in no case can $v_{\rm term}$ be smaller than $\beta c_S$, but it can
be larger. In the latter case (Case B) the contraction phase is not counted
as being a part of hot ISM, and 
\[
	q_B = {1\over 3\eta+1}.
\]

Supernova remnants affect the ambient medium in two ways: they exert
an addition pressure on the ambient gas, and, when they terminate, they
deposit the rest of their mechanical energy that was not cooled off,
into the ambient ISM. The pressure jump at the shock wave is
\[
	{\Delta P\over P_0} = {2\gamma\over\gamma+1}
	\left(\Mach^2-1\right),
\]
where $\gamma$ is the polytropic index for the gas, and $\Mach$ is the
Mach number of the shock,
\[
	\Mach = {v_{\rm term}\over c_S}.
\]
For the power-law expansion law the shock velocity changes as $t^{\eta-1}$,
and the average relative excess pressure a supernova remnant exerts on the 
ambient ISM over its lifetime is
\begin{equation}
	{\Delta P\over P_0} = {4\pi\over3\Nu}
	\int R_S(t)^3 {2\gamma\over\gamma+1}
	\left(\Mach(t)^2-1\right) dt 
	= {2\gamma\over\gamma+1} {p\over q},
	\label{dpeq}
\end{equation}
where dimensionless parameter $p$ is different for cases A and B:
\begin{description}
\item[Case A:]\ \ $\displaystyle p_A = {1\over5\eta-1} - {1\over3\eta+1}$,
\item[Case B:]\ \ $\displaystyle 
p_B = {1\over5\eta-1}\left(v_{\rm term}\over\beta c_S\right)^2 - 
{1\over3\eta+1}$.
\end{description}
Note, that in the latter case $p_B$ (and, therefore, the excess
pressure) can be very large if $v_{\rm term} \gg \beta c_S$.

After the supernova remnant dies, the rest of its energy is deposited
in the ISM as thermal heating. The total thermal energy density
deposited in the
ISM by a supernova remnant is then
\begin{equation}
	{\Delta U\over U_0} = {2\gamma\over\gamma+1} 
	\left(\Mach^2-1\right)
	\left(1+{(\gamma-1)(\Mach^2-1)\over2+(\gamma-1)\Mach^2}\right),
	\label{deeq}
\end{equation}
where $U_0$ is the energy density in the ambient medium. 

Equations (\ref{qdef}), (\ref{dpeq}), and (\ref{deeq}) can now be incorporated
in the SLH-P$^3$M code. Let the subscript 0 denotes quantities that would
be computed by the code in the absence of supernova, i.e.\ when 
$f_{\rm hot}=0$. The hydrodynamic gas pressure is corrected as
$P=P_0+\Delta P f_{\rm hot}$, the star formation rate is reduced by a factor
of $f_{\rm hot}$,
\begin{equation}
	{d\rho_*\over dt} = \left(d\rho_*\over dt\right)_0
	(1-f_{\rm hot}),
	\label{negfeed}
\end{equation}
and the thermal energy is ejected in the gas at the rate
\[
	\left(dU\over dt\right)_{SN} = 
	\Delta U f_{\rm hot} {1\over m_{SN}}
	{dm_*\over dt}.
\]
I would like to point out here that the reduction in the star formation
rate itself depends on the porosity of the IGM, which itself depends on
the star formation rate. This mutual dependence works as a powerful
feedback mechanism.

The last undefined parameter in this treatment is $m_{SN}$, the total mass
in stars formed per supernova. For a solar neighborhood (or Miller-Scalo)
IMF this value is $m_{SN}=250{\rm M}_\odot$. I introduce a new parameter,
$f_{\rm IMF}$, defined as 
\[
	f_{\rm IMF} = {250{\rm M}_\odot\over m_{SN}},
\]
to account for possible deviations from the Miller-Scalo IMF.
As argued by Silk (1996), a value of $f_{\rm IMF}\sim 4$ is required
in order to account for all metals in the clusters of galaxies.

Finally, since $p_B$ (and, therefore, $\Delta P$) increases with
increasing the termination velocity $v_{\rm term}$, one may imagine that
the effect of supernovae may be increased indefinitely by increasing
$v_{\rm term}$. However, since both $R_S$ and $t_S$ decrease with
increasing $v_{\rm term}$, the porosity of the ISM {\it decreases\/}
with increasing $v_{\rm term}$ as $Q\sim v_{\rm term}^{-19/7}$
(or $Q\sim v_{\rm term}^{-24/7}$ for $n_0<n_{\rm crit}$). Therefore,
for a sufficiently large $v_{\rm term}$, when the porosity is small,
$f_{\rm hot}=Q\sim v_{\rm term}^{-19/7}$ and the total correction to
the gas pressure (and to the thermal energy) decreases with increasing
$v_{\rm term}$ as $\Delta P\sim v_{\rm term}^{-5/7}$. There is therefore
a value for $v_{\rm term}$ when the effects of supernova on the ambient
ISM is maximal.

\section{Attachments}

Additional illustrations to this paper are available online at the
URL: \hfill\break
{\sl http://astron.berkeley.edu/\~{}gnedin/metigm.html\/}.\hfill\break
They include:
\begin{itemize}
\item color versions of Fig. \ref{figPA}-\ref{figPC};
\item color versions of Fig. \ref{figPA}-\ref{figPC} but with
the slice width of $1h^{-1}\dim{kpc}$ instead of $15h^{-1}\dim{kpc}$;
\item an MPEG video of the evolution of the object discussed in 
\S \ref{expel};
\item a color image showing the merging mechanism at work.
\end{itemize}

\end{document}